\documentclass[prb,superscriptaddress, preprintnumbers,twocolumn,showpacs]{revtex4}
\usepackage{amsmath}
\usepackage{amssymb}
\usepackage{amsthm}
\usepackage{amsfonts}
\usepackage{algorithmic}
\usepackage{enumerate}
\usepackage{latexsym}
\usepackage{graphicx}
\usepackage{bm}
\usepackage{subfigure}

\begin{document}

\title{Electron Delocalization in Gate-Tunable Gapless Silicene}
\author{Yan-Yang Zhang}
\affiliation{SKLSM, Institute of Semiconductors, Chinese Academy of
Sciences, P.O. Box 912, Beijing 100083, China} \affiliation{ICQM,
Peking University, Beijing 100871, China}
\author{Wei-Feng Tsai$^*$}
\affiliation{Department of Physics, National Sun Yat-sen University,
Kaohsiung 80424, Taiwan}
\author{Kai Chang}
\affiliation{SKLSM, Institute of Semiconductors, Chinese Academy of
Sciences, P.O. Box 912, Beijing 100083, China}
\author{X.-T. An}
\affiliation{SKLSM, Institute of Semiconductors, Chinese Academy of
Sciences, P.O. Box 912, Beijing 100083, China}
\author{G.-P. Zhang}
\affiliation{Department of Physics, Renmin University of China,
Beijing 100872, China}
\author{X.-C. Xie}
\affiliation{ICQM, Peking University, Beijing 100871, China}
\author{Shu-Shen Li}
\affiliation{SKLSM, Institute of Semiconductors, Chinese Academy of
Sciences, P.O. Box 912, Beijing 100083, China}

\date{\today}

\begin{abstract}
The application of a perpendicular electric field can drive silicene into a
gapless state, characterized by two nearly fully spin-polarized
Dirac cones owing to both relatively large spin-orbital interactions
and inversion symmetry breaking. Here we argue that since
inter-valley scattering from non-magnetic impurities is highly
suppressed by time reversal symmetry, the physics should be
effectively single-Dirac-cone like. Through numerical calculations,
we demonstrate that there is no significant backscattering from a
single impurity that is non-magnetic and unit-cell uniform,
indicating a stable delocalized state. This conjecture is then
further confirmed from a scaling of conductance for disordered
systems using the same type of impurities.

\end{abstract}

\pacs{71.23.-k, 73.21.-b, 73.43.Nq,}
\maketitle

\section{introduction}
It is well-known that a single-cone Dirac fermion is immune to backscattering and is thus hard to be localized.\cite{Ando1998,Bardarson2007,Nomura2007,CastroNeto2009} However, graphene has two Dirac cones (valleys), as required by the fermion
doubling theorem.\cite{Novoselov2004,CastroNeto2009,YJJiang2013} Consequently, in the presence of impurities, the inter-valley scattering from impurities cannot be strictly prohibited and this leads to remarkable backscattering,
resulting in localization in two dimensions (2D).\cite{Morpurgo2006,Altland2006,YYZhang2009} This is essentially
different from three-dimensional topological insulators (3DTIs),
with just one Dirac cone for each surface.\cite{Hasan2010}

Recently, silicene, which is the silicon version of graphene on a honeycomb
lattice, has been an exciting subject.
\cite{Lalmi2010,CCLiu2011,CCLiu2011B,BFeng2012} Due to its buckled
structure, the spin-orbital coupling (SOC) is highly enhanced. With
a perpendicular external electric field such structure also provides
the tunability of the bulk gap $\Delta_G$.\cite{Drummond2012} As the applied field increases, the gap closing and reopening indicates a topological
phase transition between a 2DTI and a trivial band insulator.
\cite{KaneMele,CCLiu2011,Ezawa2012,Ezawa2012B,WFTsai2013} Exactly in the
critical gapless state, where $\Delta_G=0$, the low-energy
electronic structure can be described by a massless Dirac Hamiltonian,
forming two Dirac cones. The presence of various SOC interactions on
the lattice results in rich spin textures around the Dirac points
and eventually leads to profound behaviors in response to impurity
scattering.

The most intriguing property of the gapless gated silicene, also the
focus in this work, is the opposite spin polarization at different
valleys, i.e., the valley-spin
locking.\cite{Ezawa2012,Ezawa2012B,WFTsai2013} Explicitly, the Dirac
cone around $K$ ($K'$) point is polarized with spin up (down),
mainly originating from the intrinsic SOC between next
nearest-neighbor (NNN) sites as well as broken inversion symmetry
due to the external electric field. Thus, such phase is dubbed
spin-valley-polarization metal (SVPM).\cite{Ezawa2012B} Ideally assuming no Rashba SOC, the spin around each cone is fully polarized, and,
contrary to graphene, inter-valley (also spin-flip) scattering from
non-magnetic impurities is strictly prohibited by time reversal
symmetry (TRS). Therefore, two Dirac cones in this system are
effectively decoupled and consequently the two-component,
single-flavor Dirac physics emerges. Now it is quite essential to
ask if there can be any delocalized states in the strict sense under
disorder. In addition, Rashba SOC, which includes spin-flip
processes, is nevertheless inevitable in realistic silicene. Can it
induce inter-valley scattering and lead to the breakdown of the
single Dirac cone physics as well?

To answer these questions, in this paper, we systematically study
the non-magnetic impurity scattering problem in the gapless system,
designed to capture the physics of silicene and related materials,
via numerical calculations. By comparing with various typical
arrangements of SOCs, we found that 1) from the quasi-particle
interference (QPI) pattern associated with single impurity, within a
certain region of parameter space (low energy, small Rashba SOC, and
moderate impurity scattering strength) for spin-valley-polarization
metal, the ``unit-cell impurity'' will not give rise to significant
inter- or intra-valley backscattering; 2) the positive beta function
(defined below) in the disordered system further confirms the
conclusion in 1) and suggests the existence of a truly delocalized
state.

\begin{figure*}[tb]
\includegraphics*[width=0.85\textwidth]{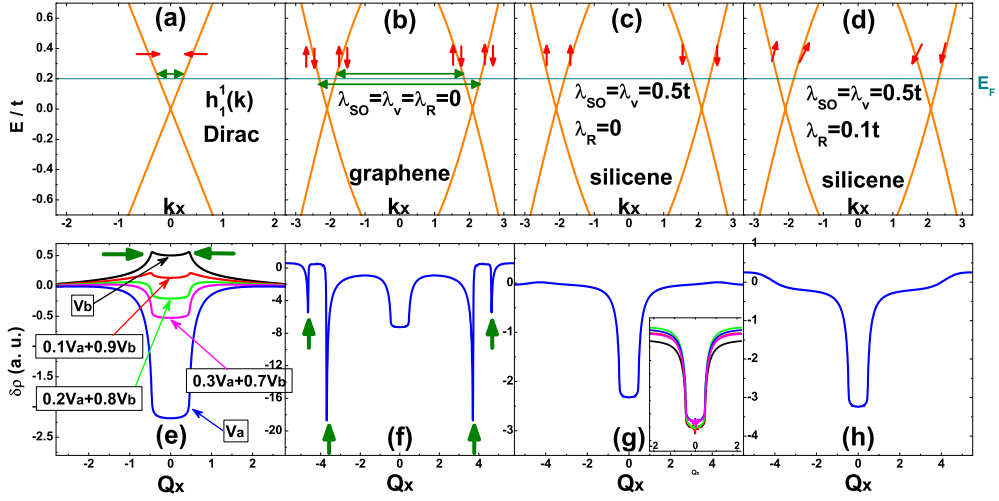}
\caption{(Color online) 2D dispersions $E(\bm{k})$ (the upper row)
and corresponding QPI curves $\delta\rho(\bm{Q})$ (the lower row)
for ideal Dirac fermion [(a) and (e)], graphene [(b) and (f)], and
silicene with either $\lambda_\mathrm{R}=0$ [(c) and (g)] or
$\lambda_\mathrm{R}=0.1t$ [(d) and (h)]. Red arrows on dispersions
illustrate the orientations of pseudo-spins (a) or physical spins
(the rest). Green thick arrows indicate significant scattering
processes. Main QPI curves are plotted along the $Q_x$-axis, while
the inset of (g) in different directions. All QPI curves are plotted
for impurity strength $V_0=t$, at Fermi energy $E_F=0.2t$, with
energy broadening $\gamma=0.005t$ and $1000\times 1000$ grid for
numerical integrations.} \label{FigMain}
\end{figure*}

\section{Hamiltonian and effective theory}
Silicene or the Ge, Sn, and Pb counterparts can be minimally described by a
tight-binding model defined on a honeycomb lattice with energy
scales $t\gg\lambda_{\mathrm{SO}}>\lambda_{\mathrm{R}}$
\cite{CCLiu2011B,Ezawa2012,Ezawa2012B},
\begin{eqnarray}
H&=&t\sum_{\langle{ij}\rangle,\sigma}c_{i\sigma}^{\dag}c_{j\sigma}
+i\frac{\lambda_{\mathrm{SO}}}{3\sqrt{3}}\sum_{\langle\langle{ij}\rangle\rangle,\sigma\sigma^\prime}
\nu_{ij}c_{i\sigma}^{\dag}s^z_{\sigma\sigma^\prime}c_{j\sigma^\prime}\nonumber\\
&-& i\frac{2\lambda_{\mathrm{R}}}{3}\sum_{\langle\langle{ij}\rangle\rangle,\sigma\sigma^\prime}
\mu_{ij}c_{i\sigma}^{\dag}(\bm{s}\times\bm{\hat{d}}_{ij})^z_{\sigma\sigma^\prime}c_{j\sigma^\prime}\nonumber\\
&+&\lambda_{\nu}\sum_{i,\sigma}\xi_{i}c_{i\sigma}^{\dag}c_{i\sigma}.
\label{eq1}
\end{eqnarray}
The first term describes the nearest-neighbor (NN) hopping, where $c_{i\sigma}^{\dag}$ creates an
electron at site $i$ with spin polarization $\sigma$. The second
term represents the intrinsic SOC between NNN sites, where
$\bm{s}=(s_x,s_y,s_z)$ are the Pauli matrices for physical spins,
and $\nu_{ij}=(\bm{d}_i\times \bm{d}_j)_z/|\bm{d}_i\times
\bm{d}_j|=\pm 1$ with $\bm{d}_i$ and $\bm{d}_j$ the two NN bonds
connecting NNN sites $i$ and $j$. The third term is the NNN Rashba
SOC,\cite{NNRashba} where $\mu_{ij}=\pm 1$ for the A and B sites,
respectively, and $\bm{\hat{d}}_{ij}=\bm{d}_{ij}/|\bm{d}_{ij}|$
represent the unit vector of $\bm{d}_{ij}$ which connects NNN
sites $i$ and $j$. The fourth term represents the staggered
potential, and the strength $\lambda_\mathrm{v}=l_z E_z$ can be
tuned by a perpendicular electric field $E_z$ because of the
buckling distance $l_z$ between two sublattices. The model parameters
for silicene are $t=1.04$eV, $\lambda_\mathrm{SO}=4.2$meV, $\lambda_\mathrm{R}=8.66$meV
and $l_z=0.035$e\AA\cite{WFTsai2013}. Note that if we only keep the
first term with $t=2.7$eV, Eq. (\ref{eq1}) simply describes undoped
graphene\cite{CastroNeto2009}. Hereafter, we adopt $t$ as the energy unit and lattice
constant $a$ (NNN distance) as the length unit.

Around two Dirac points at $K(K')=(\pm4\pi/3,0)$ in $k$-space, the
low-energy effective Hamiltonian for Eq. (\ref{eq1}) with the basis
$(\psi_{A\uparrow},\psi_{B\uparrow},\psi_{A\downarrow},\psi_{B\downarrow})^{T}$
reads
\begin{eqnarray}
H^{\eta}_1(\bm{q}) &=&\left(
\begin{array}{cc}
h^{\eta}_1(\bm{q}) & g_1(\bm{q}) \\
g_1^{\dagger }(\bm{q}) & h^{\eta}_1(\bm{q})%
\end{array}%
\right),  \label{eqH} \\
h^{\eta}_1(\bm{q}) &=&\left(
\begin{array}{cc}
-\eta\lambda_{\mathrm{SO}}+\lambda_{\mathrm{v}} & \frac{\sqrt{3}}{2}t(-\eta q_x-iq_y) \\
\frac{\sqrt{3}}{2}t(-\eta q_x+iq_y) & \eta\lambda_{\mathrm{SO}}-\lambda_{\mathrm{v}}%
\end{array}%
\right),  \label{eqh1} \\
g_1(\bm{q}) &=&\left(
\begin{array}{cc}
\lambda_{\mathrm{R}}(iq_x+q_y) & 0 \\
0 & -\lambda_{\mathrm{R}}(iq_x+q_y)%
\end{array}%
\right),  \label{eqg} \\
  \notag
\end{eqnarray}
where $\bm{q}$ is measured from the Dirac point, $\eta=\pm1$ for $K$
($K'$) point is the valley index, and $h^{\eta}_1(\bm{q})$ is just
the ideal Dirac fermion Hamiltonian for pseudo-spin, with Fermi
velocity $\frac{\sqrt{3}}{2}t$ and mass
$\lambda_{\mathrm{v}}-\eta\lambda_{\mathrm{SO}}$. Gating the system
such that $\lambda_\mathrm{v}=\lambda_{\mathrm{SO}}$ but with
$\lambda_{\mathrm{R}}=0$, the full spin polarization of the valleys
can be clearly seen: At valley $K$, the spin-up bands are {\it
gapless} forming a Dirac cone, in contrast to spin-down bands now
separated by a gap
$\Delta_G=2|\lambda_{\mathrm{SO}}+\lambda_{\mathrm{v}}|$ and thus
out of the low-energy regime; at valley $K'$ it is in opposite
orientation due to TRS. The presence of considerable
$\lambda_{\mathrm{R}}$ destroys this full spin polarization but the
majority around each valley does not change. Such states with two
massless Dirac cones will be the main focus throughout this work.
Restricting $\lambda_{\mathrm{v}}=\lambda_{\mathrm{SO}}$ (therefore
$\Delta_G=0$) while allowing one to vary their strengths as well as the
values of the Fermi level and $\lambda_{\mathrm{R}}$ in the system
give rise to rich physics, which reflects the interplay among spin,
sublattice (pseudo-spin), and valley degrees of freedom under
non-magnetic impurity scattering.

\section{QPI from single impurity}
The focus of the current study will be the effect of electronic scattering from impurity {\it potentials}.
Impurity potentials can be induced by atomic substitution, surface adsorption or by the substrate
under the 2D sample. Among various origins, the adsorption of different atoms for silicene has been
discussed from {\it ab initio} calculations recently\cite{RGQuhe2012,XQLin2012,HSahin2013,JSivek2013}.
In particular, it has been found that silicene tends to adsorb adatoms (including metal atoms) more strongly than
graphene.
Depending on which element is concerned, the adatom can sit on
the ``hill'', ``valley'', ``bridge ''or ``hollow'' positions of the hexagonal ring respectively\cite{HSahin2013,JSivek2013}.

We first investigate the scattering from a single impurity, by calculating QPI pattern.\cite{QHWang2003} The Green's function for the clean system is
$G^0(E,\bm{k})\equiv
G^0(E,\bm{k},\bm{k})=[(E+i\gamma)I-H(\bm{k})]^{-1}$, where $I$ is
the identity matrix and $\gamma\ll 1$ is the energy broadening. Here
we only consider a single impurity with potential
$\sim\delta(\bm{x})$ in a definite unit cell so that the impurity
matrix $V(\bm{k}_1,\bm{k}_2)=V$ is independent of $\bm{k}$. The
impurity induced Green's function is expressed as
\begin{equation}
\delta
G(E,\bm{k}_1,\bm{k}_2)=G^0(E,\bm{k}_1)T(E,\bm{k}_1,\bm{k}_2)G^0(E,\bm{k}_2).
\label{eq5}
\end{equation}
The standard perturbation method gives \cite{QHWang2003}
\begin{equation}
T(E)=[I-V\Gamma^0(E)]^{-1}V,\label{eq5b}
\end{equation}
where $\Gamma^0(E)=\int\frac{d^2k}{(2\pi)^2}G^0(\bm{k},E)$. Now the
Fourier transform of the induced local density of states is
\begin{equation}
\delta\rho(E,\bm{Q})=\frac{i}{2\pi}\int\frac{d^2k}{(2\pi)^2}g(E,\bm{k},\bm{Q}),
\label{eq7}
\end{equation}
where $\bm{Q}=\bm{k}'-\bm{k}$ and
$g(E,\bm{k},\bm{Q})=\mathrm{Tr}\big(\delta
G(E,\bm{k},\bm{k}')-\delta G^*(E,\bm{k}',\bm{k})\big)$. The spectrum
$\delta\rho(E,\bm{Q})$ in Eq.~(\ref{eq7}) is called the QPI pattern,
which can also be obtained experimentally from the Fourier
transformation of scanning tunneling microscopy (STM) measurements.\cite{Hoffman2002,Roushan2009} This pattern provides an intuitive picture of scattering processes: Significant scattering processes will manifest themselves as peaks in the QPI pattern with associated scattering momenta $\bm{Q}$.

\begin{figure}[tb]
\includegraphics*[width=0.50\textwidth]{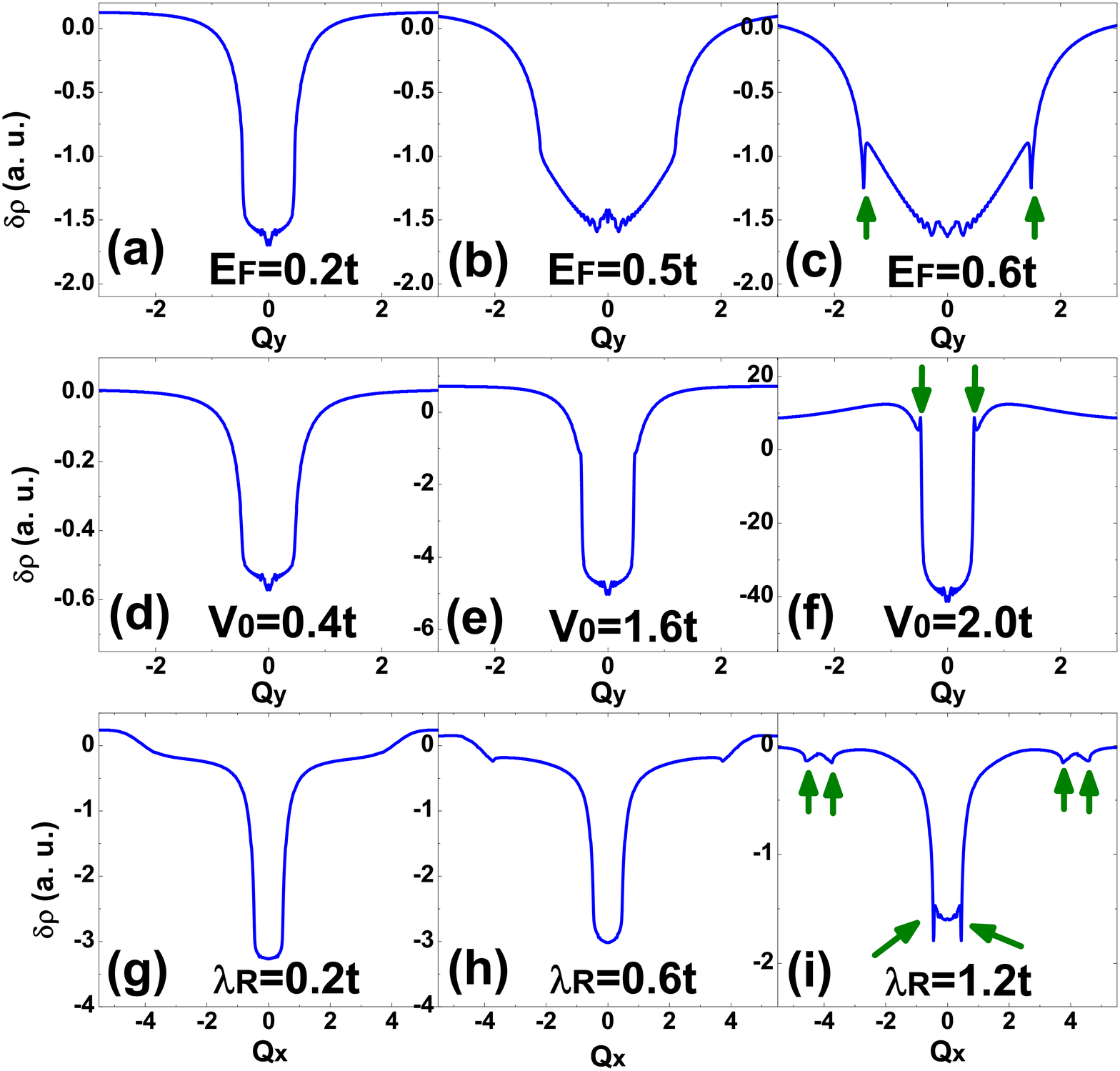}
\caption{(Color online) QPI pattern for silicene with different
Fermi energies $E_F$, impurity strengths $V_0$, and Rashba SOC
$\lambda_{\mathrm{R}}$. The upper row (a-c): $V_0=t$ and
$\lambda_{\mathrm{R}}=0$; The middle row (d-f): $E_F=0.2t$ and
$\lambda_{\mathrm{R}}=0$; The lower row (g-i): $E_F=0.2t$ and
$V_0=t$. The scanning angle is chosen along the $Q_y$ axis from (a)
to (f), where possible intra-valley backscattering reaches its
maximum amplitude \cite{Ando1998}, and along $Q_x$ axis from (g) to
(i) for the detection of possible inter-valley scattering. All other
parameters are the same with Fig.~\ref{FigMain}(g).} \label{FigEfW}
\end{figure}

As a warm-up but essential example, we start with the single valley,
single spin, $2\times 2$ ideal Dirac fermion Hamiltonian with just
linear terms,
$h^{\eta=1}_1(\bm{k})=\frac{\sqrt{3}t}{2}(-k_x\tau_x+k_y\tau_y)$
(see Eq. (\ref{eqh1})), with $\tau_i$ the Pauli matrix acting on
sublattice (pseudo-spin) space. The impurity potential in $k$-space
is diagonal as $V_a\equiv V_0\tau_0$, $V_b\equiv V_0\tau_3$, or
their combinations with relative weight $r$,
\begin{equation}
V=r\cdot V_a+(1-r)\cdot V_b,\qquad 0\leq r\leq 1
\end{equation}
The computed QPI, $\delta\rho(\bm{Q})$ of $V_a$, is plotted as the
blue curve in Fig.~\ref{FigMain}(e). The curve has no significant
scattering peaks, consistent with the well-known fact that $V_a$
cannot induce backscattering for a massless ideal Dirac
fermion.\cite{Ando1998,Bardarson2007,Nomura2007} Notice that in the
language of sublattice as pseudo-spin $\tau$, $V_a$ corresponds to a
``unit-cell impurity'' which is uniform within two sites of a unit
cell. On the other hand, we also show the QPI for the ``site
impurity'', $V_b$, in Fig.~\ref{FigMain}(e) as the black curve. Two
peaks associated with intra-valley backscattering can be seen. This
is not surprising because $V_b$ is a mass term for ideal Dirac
fermion and destroys the pseudo-``TRS'', leading to a tendency
towards localization.\cite{Ando1998} From results for an impurity
with different weights of $V_a$ and $V_b$ also shown in
Fig.~\ref{FigMain}(e), it is interesting to notice that, a small
weight ($r\gtrsim 20\%$) of $V_a$ is sufficient to annihilate the
significant backscattering peaks into the smooth background. In real
space, an impurity with finite $V_a$ component corresponds to a long
range one, with a smooth potential configuration within the unit
cell. Such impurities can be dominant in
graphene\cite{CastroNeto2009} and therefore should also be easily
realized experimentally for silicene. Moreover, the ``hollow-type''
and ``bridge-type'' adsorptions in
silicene\cite{HSahin2013,JSivek2013}, which do not induce strong
staggered potential, should also play such a role. In the rest of
the paper, we will restrict ourselves to the discussions of
unit-cell impurity $V_a$ .

In the following, we will consider the full tight-binding $4\times4$
Hamiltonian $H(\bm{k})$, i.e., the $k-$representation of $H$ in
Eq.~(\ref{eq1}). For comparison purpose, we first take parameters
$\lambda_{\mathrm{SO}}=\lambda_{\mathrm{v}}=\lambda_{\mathrm{R}}=0$
such that $H(\bm{k})$ describes graphene without SOC. Different from
the case with ideal Dirac fermion Hamiltonian, $H(\bm{k})$ here has
two important features: The existence of two spins and two valleys, and
higher order corrections (trigonal warping) within each valley, as
illustrated in Fig.~\ref{FigMain}(b). Given a unit-cell impurity
potential,
\begin{equation}
V_a=V_0 s_0\otimes\tau_0=\mathrm{diag}\big(V_0,V_0,V_0,V_0\big),\label{eqVa4}
\end{equation}
which is ``non-magnetic'' both for physical spin $s$ and pseudo-spin
$\tau$, the corresponding QPI is shown in Fig.~\ref{FigMain}(f). It
has very sharp peaks associated with inter-valley backscatterings
between states with opposite $\bm{k}$ and velocity, as indicated by
the green arrows. As in ordinary orthogonal disordered systems in 2D,
\cite{Anderson1958,PALee1985} this strong backscattering is
responsible for the localization in graphene
\cite{Morpurgo2006,Altland2006,YYZhang2009} and weak 3DTI.\cite{Hasan2010} In short, the coupling between two Dirac cones
(with opposite Berry curvatures \cite{DXiao2007}) makes the physics
rather trivial.

Armed with QPI studies from above two examples, we come to our main
focus, gapless silicene with
$\lambda_{\mathrm{SO}}=\lambda_{\mathrm{v}}=0.5t$ for $H(\bm{k})$.
Such large SOC is taken simply for giving enough space to extract
out clear physics within our numerical precisions. No qualitative
difference is expected as long as Fermi energy
$E_F\in\big(-|\lambda_{\mathrm{SO}}+\lambda_{\mathrm{v}}|,|\lambda_{\mathrm{SO}}+\lambda_{\mathrm{v}}|\big)$,
where the SPVM picture holds.

\begin{figure*}[tb]
\includegraphics*[bb=80 0 970 345, width=0.75\textwidth]{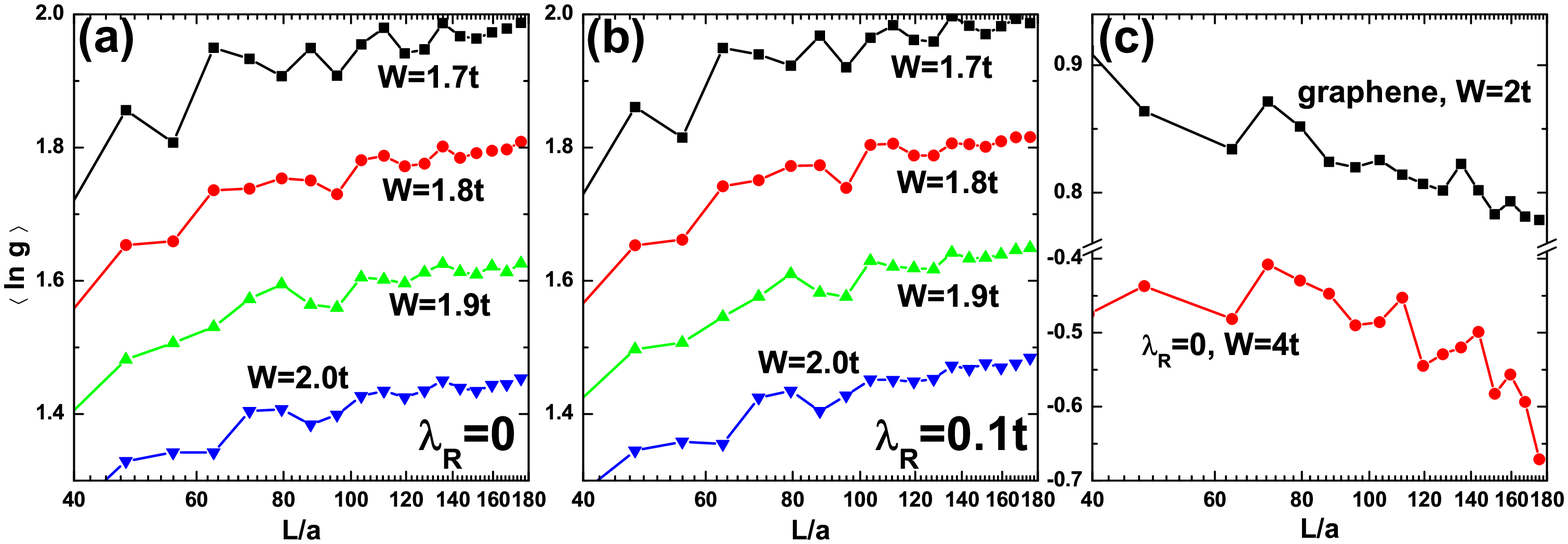}
\caption{(Color online) Typical conductance as a function of the
system size (in logarithmic scale). (a)
$\lambda_\mathrm{SO}=\lambda_\mathrm{v}=0.5t$,
$\lambda_\mathrm{R}=0$; (b) $\lambda_\mathrm{R}=0.1t$; (c) Two
examples of localization: graphene with $W=2t$ (black), and silicene
with $\lambda_\mathrm{SO}=\lambda_\mathrm{v}=0.5t$,
$\lambda_\mathrm{R}=0$, under strong
disorder $W=4t$ (red). Each dot is the average over 2000 disorder
samples with unit cell impurities.} \label{FigG}
\end{figure*}

We first consider the spin conserved case $\lambda_{\mathrm{R}}=0$,
where each valley is fully spin polarized [See
Fig.~\ref{FigMain}(c)], therefore inter-valley scattering is
prohibited. Indeed, with the same $V_a$ in Eq.~(\ref{eqVa4}), now
the QPI in Fig.~\ref{FigMain}(g) is qualitatively different from
that of graphene, but similar to that of the ideal Dirac fermion
[Fig.~\ref{FigMain}(e)]. Moreover, there are no significant
\emph{intra}-valley backscattering peaks, either. The intra-valley
features are almost isotropic in $\bm{Q}$, as shown in the inset of
Fig.~\ref{FigMain}(g), even though the full band structure is
anisotropic around each Dirac point. It was argued that trigonal
warping would lead to nonzero backscattering
amplitude.\cite{Ando1998} However, our numerical results show that
such backscattering is very weak and could be immersed in the
continuum background of other scattering processes, reflected by the
absence of associated distinguishable peaks. Therefore, the gapless
silicene with broken inversion symmetry effectively exhibits the
massless Dirac fermion physics, so long as the Fermi energy is not
far from the Dirac point and the impurity strength is not strong
(i.e., $|E_F|+|V_0| < E$, where $E\sim
O(\lambda_\mathrm{SO}+\lambda_\mathrm{v})$, the energy scale which
protects the spin-valley-polarization metal phase). This is one of
the important findings in this work. The absence of remarkable
backscattering should signify a delocalized state to disorders, as
will be numerically verified later.

Before entering into the discussion on disordered systems, two
remarks are in order. First, in Figs.~\ref{FigMain}(c) and (g), with
vanishing $\lambda_\mathrm{R}$, inter-valley scattering is in fact
suppressed {\it a priori}. Nonzero $\lambda_\mathrm{R}$, as to be the
case in silicene, makes the spin-valley polarization imperfect [See
Fig.~\ref{FigMain}(d)]. However, as shown in Fig.~\ref{FigMain}(h),
it is remarkable to see that such Rashba term does not give rise to
a significant inter-valley scattering, and thus the effective ``single-valley Dirac physics'' remains intact. This can be due to the following two intuitive reasons: 1) The NNN Rashba interaction makes no contribution at $K$ ($K'$) points and thus its effect is also expected to be small around $K$ ($K'$) points; 2) the full backward scattering, which relates inter-valley points, {\it i.e.}, time-reversal partners (still with opposite spin polarizations under
Rashba interaction), is difficult to happen through {\it nonmagnetic}
impurity.

Second, further
increasing a parameter such as $E_F$, impurity strength $V_0$, or
$\lambda_{\mathrm{R}}$ in the system is expected to enhance intra-
and inter-valley scattering processes due to unavoidable
contributions from higher order corrections and spin/valley mixing.
Indeed, as clearly shown in Fig.~\ref{FigEfW}, the QPI pattern
changes at some point, indicating a transition from a delocalized to
localized state beyond effective single-valley Dirac physics. For
instance, in the case of very strong $\lambda_\mathrm{R}$ in
Fig.~\ref{FigEfW} (i), although two states $\big|\bm{k}\big\rangle$
and $\big|-\bm{k}\big\rangle$ in different valleys (with exactly
opposite spin orientations) cannot be coupled by a non-magnetic
impurity, the spin orientations in their neighborhoods will not be
exactly opposite. Thus an inter-valley backscattering can be allowed
due to the energy broadening $\gamma$.

\section{Scaling of conductance: multiple impurities}
So far, the scattering from a single impurity has been investigated. If the
backscattering is effectively ignorable, does this really lead to
delocalized state in disordered gapless silicene with unit-cell
impurities? To confirm that it does, we perform a standard numerical scaling
for disordered silicene. Disorder is added to the Hamiltonian
(\ref{eq1}) as $\sum_{i,\sigma}\epsilon_i
c_{i\sigma}^{\dag}c_{i\sigma}$, where $\epsilon_i$ is a random
number uniformly distributed within $\big(-W/2,W/2\big)$. Here
$\epsilon_i$ is independent of spin due to TRS. If $\epsilon_i$ is
further identical for two sites in each unit cell, then it
corresponds to unit cell impurities $V_a$. In realistic silicene material,
such impurities can be long range impurities as in graphene\cite{CastroNeto2009}, or the
``hollow'' and ``bridge'' types of adsorbed impurities as reported in Refs. \onlinecite{HSahin2013} and \onlinecite{JSivek2013}.
The intrinsic conductance $g$ is defined as $1/g=1/g_L-1/N_c$, where $g_L$ is the two-terminal quantum conductance, $N_c$ is the number of propagating channels and $1/N_c$ is the contact resistance.\cite{Braun1997} This $g$ is
suitable for a numerical scaling \cite{Slevin2001,YYZhang2009}
\begin{equation}
\beta=\frac{d\langle \ln g\rangle}{d\ln L},\label{eq11}
\end{equation}
where $\langle \cdots\rangle$ is the average over random ensemble,
and $L$ is the spatial size of the sample with a fixed ratio of
length and width. This scaling function $\beta$ is used as a
criteria: $\beta<0$ and $\beta>0$ correspond to localized and
delocalized states, respectively.

In Fig.~\ref{FigG}, we plot $\langle \ln g \rangle$ as a function of
size $L$ (in logarithmic scale), where the slope represents $\beta$.
It can be seen from Figs.~\ref{FigG}(a) and (b) that, for unit-cell
impurities, apart from some fluctuations due to the smallness of
conducting channels, $\langle \ln g\rangle $ is clearly increasing
with increasing $L$, suggesting a delocalized state with $\beta>0$.
These are consistent with our results of the absence of significant
backscattering from the single impurity study, further confirming
the robustness of the effective single-valley Dirac physics. Note
that this is totally different from the case of graphene with the
$W=2t$ [black curve in Fig.~\ref{FigG}(c)], where the slope is
negative. It has been found that for graphene, even long-range
impurities cannot  maintain a fully delocalized state with $\beta>0$
because of inevitable inter-valley scattering.\cite{YYZhang2009}
Of course, as in any lattice models, sufficiently strong disorder,
for instance with $V_0\gg O(\lambda_{\mathrm{SO}}+\lambda_{\mathrm{v}})$,
will eventually localize all the electrons, as the red curve in Fig. \ref{FigG}(c) shows.
Therefore, it is natural to expect rich
localization-delocalization transition behavior in the parameter
space spanned by $E_F$, $W$, $\lambda_{\mathrm{R}}$, and
$\lambda_{\mathrm{SO}}$ $(=\lambda_{\mathrm{v}})$. More details of
such localization-delocalization transition, e.g., the universality,
critical exponents, and global phase diagram will be discussed
elsewhere.

Delocalized bulk states in the doped Kane-Mele model with nontrivial
Z$_2$ topological nature were found in Ref.~\onlinecite{Onoda2007}.
In that case, delocalized states can only appear when large (comparable to $E_F$) and inversion-broken NN Rashba SOC is nonzero, making the system truly {\it symplectic}. Otherwise, the system is just decoupled into two gapped {\it unitary} subsystems, namely, two massive Dirac cones around $K$ and $K'$, where no states with $\beta>0$ can be observed.\cite{Onoda2007} This is indeed reasonable as a gapped Dirac cone has serious backscattering.\cite{Ando1998} In our case,
however, the physics behind delocalization lies on either independent {\it unitary} subsystems, {each of which owns a massless Dirac cone} (zero NNN Rashba SOC), or {\it symplectic} subsystems with two nearly independent gapless Dirac cones (nonzero NNN Rashba SOC). The Dirac cones are already non-degenerate with almost fully spin polarization (along the $z$ axis) as long as the  inversion-symmetric, NNN $\lambda_{\mathrm{R}}$ interaction is small.

\section{Discussion and conclusion}
The essence of this work is to reveal the physics behind the delocalization phenomenon, which can be understood from the picture of effectively decoupled gapless Dirac cones. This picture is valid in certain parameter ranges. Here we emphasize again the relevant energy scales important to any experimental demonstration for the delocalization in silicene, such as the presence of a robust Dirac point and linear dispersion in STM or the weak antilocalization in the magnetoresistance measurement.
First, the gapless condition
$\lambda_\mathrm{v}=l_z E_z=\lambda_\mathrm{SO}$ gives the critical electric field $E_z=\lambda_\mathrm{SO}/l_z\sim 0.12$V\AA$^{-1}$, which is experimentally achievable. In this case, the half width of the energy window for spin-valley locking is
$|\lambda_{\mathrm{SO}}+\lambda_{\mathrm{v}}|=2|\lambda_{\mathrm{SO}}|\sim 8.4$ meV. This range can be even larger in the Ge, Sn or Pb counterpart.\cite{WFTsai2013}
On the other hand, such energy window is still small enough to keep
the dispersion of Dirac fermions linear. Although NNN Rashba SOC breaks the perfect spin-valley locking, $\lambda_{\mathrm{R}}=8.66$ meV is less than 1\% of $t$, and therefore this effect is very weak.

In summary, we reveal the essential transport properties via
numerical simulations on a critically gated buckled honeycomb
structure of silicene (and also suitable for the Ge, Sn, and Pb
counterparts) under non-magnetic impurity scattering. In particular, as long as $|E_F|+|V_0| < E$ with $E$, an energy scale of the order of $\lambda_\mathrm{SO}+\lambda_\mathrm{v}$, we find: 1) QPI by a single unit-cell impurity shows no significant backscattering, suggesting an effective {\it single-valley} Dirac physics, in spite of
weak trigonal warping. 2) The robustness of such delocalized state is
further confirmed by the positiveness of the $\beta$ function for a
disordered system, even in the presence of Rashba SOC. Our finding sheds a new light on constructing high mobility silicene-based
electronic devices. Moreover, we believe our result is also insightful to
relevant systems such as a 2D MoS$_{2}$ \cite{DXiao2012} and a
cold-atom system with arranged SOC.\cite{Goldman2009}

\begin{acknowledgements}
YYZ thanks Q. F. Sun, H. Jiang and H. W. Liu for beneficial
discussions. This work was supported by NSFC (Grant No. 11204294)
and 973 Program Project No. 2013CB933304. WFT is supported by the
NSC in Taiwan under Grant No. 102-2112-M-110-009.
\end{acknowledgements}


\begin{thebibliography}{99}

\bibitem{Email} *To whom correspondence should be addressed. Email: wftsai@mail.nsysu.edu.tw

\bibitem{CastroNeto2009} A. H. Castro Neto, F. Guinea, N. M. R. Peres,
K. S. Novoselov and A. K. Geim, Rev. Mod. Phys. \textbf{81}, 109
(2009).

\bibitem{Ando1998} T. Ando, T. Nakanishi and R. Saito, J. Phys. Soc. Jpn. \textbf{67},
2857 (1998).

\bibitem{Bardarson2007} J. H. Bardarson, J. Tworzyd{\l}o, P. W. Brouwer and C. W. J. Beenakker,
Phys. Rev. Lett. \textbf{99}, 146806 (2007).

\bibitem{Nomura2007} K. Nomura, M. Koshino and S. Ryu,
Phys. Rev. Lett. \textbf{99}, 106801 (2007).

\bibitem{Novoselov2004} K. S. Novoselov, A. K. Geim, S. V. Morozov,
D. Jiang, Y. Zhang, S. V. Dubonos, I. V. Grigorieva, A. A. Firsov,
Science \textbf{306}, 666 (2004).

\bibitem{YJJiang2013} Y.-J. Jiang, T. Low, K. Chang, M. I. Katsnelson and F. Guin,
Phys. Rev. Lett. \textbf{110}, 046601 (2013).

\bibitem{Morpurgo2006} A. F. Morpurgo and F. Guinea, Phys. Rev. Lett. \textbf{97},
196804 (2006).

\bibitem{Altland2006} A. Altland, Phys. Rev. Lett. \textbf{97},
236802 (2006).

\bibitem{YYZhang2009} Y.-Y. Zhang, J.-P. Hu, B. A. Bernevig, X. R. Wang,
X. C. Xie and W. M. Liu, Phys. Rev. Lett. \textbf{102}, 106401
(2009).

\bibitem{Hasan2010} See, for instance, M. Z. Hasan and C. L. Kane, Rev. Mod. Phys. \textbf{82}, 3045 (2010) and reference therein.

\bibitem{Lalmi2010} B. Lalmi, H. Oughaddou, H. Enriquez, A. Kara, S. Vizzini, B. Ealet and
B. Aufray, Appl. Phys. Lett. \textbf{97}, 223109 (2010).

\bibitem{CCLiu2011} C.-C. Liu, W. Feng and Y. Yao, Phys. Rev. Lett. \textbf{107}, 076802
(2011).

\bibitem{BFeng2012} B. Feng,Z. Ding, S. Meng,Y. Yao, X. He, P. Cheng, L. Chen
and K. Wu, Nano Lett. \textbf{12}, 3507 (2012).

\bibitem{CCLiu2011B} C.-C. Liu, H. Jiang and Y. Yao, Phys. Rev. B \textbf{84},
195430 (2011).

\bibitem{KaneMele} C. L. Kane and E. J. Mele, Phys. Rev. Lett. \textbf{95}, 146802
(2005); 226801 (2005).

\bibitem{Drummond2012} N. D. Drummond, V. Zolyomi and V. I. Fal'ko, Phys. Rev. B \textbf{85}, 075423 (2012).

\bibitem{Ezawa2012} M. Ezawa, New J. Phys. \textbf{14},
033003 (2012).

\bibitem{Ezawa2012B} M. Ezawa, Phys. Rev. Lett. \textbf{109},
055502 (2012).

\bibitem{WFTsai2013} W.-F. Tsai, C.-Y Huang, T.-R Chang, H. Lin, H.-T. Jeng, and A. Bansil, Nat. Commun. \textbf{4}, 1500 (2013).

\bibitem{NNRashba} An NN Rashba SOC term could also be induced by inversion symmetry breaking in the gated silicene.
 However, such coupling is estimated to be much smaller than $\lambda_R$ and becomes zero at our focused gapless state.\cite{Ezawa2012B} Thus, one can safely ignore it.

 \bibitem{RGQuhe2012} R. Quhe, R. Fei, Q. Liu, J. Zheng, H. Li, C. Xu, Z. Ni,
Y. Wang, D. Yu, Z. Gao, and Jing Lu, Sci. Rep. \textbf{2},
853 (2012).

\bibitem{XQLin2012} X. Lin and J. Ni, Phys. Rev. B \textbf{86}, 075440 (2012).

\bibitem{HSahin2013} H. Sahin and F. M. Peeters, Phys. Rev. B \textbf{87}, 085423 (2013).

\bibitem{JSivek2013} J. Sivek, H. Sahin, B. Partoens, and F. M. Peeters, Phys. Rev. B \textbf{87}, 085444 (2013).

\bibitem{QHWang2003} Q.-H. Wang and D.-H. Lee, Phys. Rev. B \textbf{67},
020511(R) (2003).

\bibitem{Hoffman2002} J. E. Hoffman, K. McElroy, D.-H. Lee, K. M Lang, H. Eisaki, S.
Uchida, J. C. Davis, Science \textbf{297}, 1148 (2002).

\bibitem{Roushan2009} P. Roushan, J. Seo, C. V. Parker, Y. S. Hor, D. Hsieh, D. Qian, A.
Richardella, M. Z. Hasan, R. J. Cava, and Ali Yazdani, Nature
(London) \textbf{460}, 1106 (2009).

\bibitem{Onoda2007} M. Onoda, Y. Avishai and N. Nagaosa, Phys. Rev. Lett. \textbf{98}, 076802 (2007).

\bibitem{Anderson1958} P. W. Anderson, Phys. Rev. \textbf{109}, 1492 (1958).

\bibitem{PALee1985} P. A. Lee and T.V. Ramakrishnan, Rev. Mod. Phys. \textbf{57}, 287 (1985).

\bibitem{DXiao2007} D. Xiao, W. Yao and Q. Niu, Phys. Rev. Lett. \textbf{99}, 236809
(2007).

\bibitem{Braun1997} D. Braun, E. Hofstetter, A. MacKinnon and G. Montambaux, Phys. Rev. B \textbf{55}, 7557 (1997).

\bibitem{Slevin2001} K. Slevin, P. Marko\v{s} and T. Ohstuki, Phys. Rev. Lett. \textbf{86}, 3594
(2001).

\bibitem{DXiao2012} D. Xiao, G.-B. Liu, W.-X. Feng, X.-D. Xu and W. Yao, Phys. Rev. Lett. \textbf{108},
196802 (2012).

\bibitem{Goldman2009} N. Goldman, A. Kubasiak, A. Bermudez, P. Gaspard, M. Lewenstein
and M. A. Martin-Delgado, Phys. Rev. Lett. \textbf{103}, 035301
(2009).

\end{thebibliography}
\end{document}